\documentclass[]{aa}

\usepackage{graphicx}
\newcommand{\be}{\begin{eqnarray}}
\newcommand{\ee}{\end{eqnarray}}

\begin{document}

\title{Parameter Study of Star-Discs Encounters}
\author{S. Pfalzner \and P. Vogel \and J. Scharw\"achter \and C. Olczak}
\institute{I. Physikalisches Institut, University of Cologne, Germany}
\date{Received   / Accepted   }

\abstract{Interactions between disc-surrounded stars might play a vital 
role in the formation of planetary systems. Here a first parameter study of 
the effects of encounters on low-mass discs is presented. 
The dependence of the mass and angular momentum transport on the periastron 
distance, the relative mass of the encountering stars and eccentricity of the 
encounter is investigated in detail. This is done for prograde and 
retrograde coplanar encounters as well as non-coplanar encounters. 
For distant coplanar encounters our simulation results agree with 
the analytical approximation of the angular momentum loss by Ostriker(1994). 
However, for close or high-mass encounters, significant differences 
to this approximation are
found. This is especially so in the case of retrograde encounters, where 
the analytical result predict no angular momentum loss regardless of the 
periastron distance whereas the simulations find up to $\sim$ 20\% loss 
for close encounters. For the non-coplanar case a more complex dependency on 
the inclination between orbital path and disc plane is found than for distant 
encounters. For the coplanar prograde case new fitting formulae 
for the mass and angular momentum loss are obtained, which cover the whole
range from grazing to distant encounters. In addition, the final disc 
size and the mass exchange between discs is examined, demonstrating that 
for equal mass stars in encounters as close as 1.5 the
disc radius,  the disc size only is reduced by approximately 10\%.
}

\maketitle

\section{Introduction}

The discovery of planets outside the solar system and detailed 
observations of protoplanetary discs have been milestones in the understanding 
of planetary systems; findings which have caused long-standing theories about 
the formation of planets to be questioned. Currently there are a number of 
scenarios proposed to explain the formation of planets from the 
protoplanetary accretion disc surrounding young stars, of which two different 
schemes - planet formation by the agglomeration of matter in the disc and 
planet formation induced by gravitational disc instabilities - are the most 
favoured. In other scenarios, magnetic fields 
(\cite{balbus:apj02,klahr:apj03}) are essential. 
In several of these suggested models, encounters between stars surrounded
by their protoplanetary accretion discs play an important role
(\cite{bonnell:mnras01,kobayashi:01,bate:mnras02,oxley:mnras04}). 

As star-disc systems have a much larger cross-section than stars alone, 
interactions between two such systems are much more likely than star-star 
collisions. The discs are influenced by the encountered system even if the
periastron $r_{peri}$ of the encounter is several times larger than the 
disc size $r_{disc}$. Although distant encounters are more likely than close 
encounters, close encounters are not an unlikely event. For example,
simulations show that in the ONC cluster around 20\% of all stars undergo
encounters closer than 300AU within 3 Myr (\cite{olczak:05}).

There have been attempts to treat the problem of the encounter of star-disc 
systems analytically (\cite{ostriker:apj94}), but for close encounters the 
problem becomes highly non-linear and requires a numerical treatment.
In the 1990s several authors investigated the effect on star-disc systems
which are disturbed either by an other star or star-discs systems using 
numerical methods
(\cite{heller:apj95,hall:mnras96,hall:mnras97,larwood:97,boffin:mnras98}). At that time 
the question of binary formation was the driving force behind these
investigations. This meant only parabolic encounters were of interest
and only a limited number of cases were investigated due to the computational 
expense of such investigations. Of these studies the work by Hall et al.(1996)
spanned the widest parameter range; they treated parabolic encounters
of equal mass stars but with a preference of penetrating encounters.
More recent investigations (Pfalzner 2003, 2004a) have included hyperbolic 
encounters but concentrated on coplanar prograde encounters only.  
A comprehensive data base for all type of disc encounters - whether
prograde or retrograde, parabolic or hyperbolic, coplanar or non-coplanar - 
is still lacking.

This paper provides a first attempt at such a data base, describing the effect 
of encounters on the global disc features like mass and angular momentum loss,
change in disc size etc. for low-mass discs. Due to the computational expense, 
the parameter space is still limited and an extension will be necessary in 
the future.  

Like in most previous studies, only one of the stars involved in the
encounter is surrounded by a disc. Although this was initially done
to reduce the complexity and parameter space, for the case
of low-mass discs it turns out that the results of star-disc encounters can 
be relatively easily generalized to disc-disc encounters 
(Pfalzner et al. 2004b), making separate studies unnecessary. It should be 
pointed out that this is not so for high-mass discs.

In order to keep the computational effort as small as possible while
maximizing the accessible parameter space, in our simulations only the forces 
of the disc onto the star and vice versa are considered. The effects of 
pressure, viscosity and self-gravity within the disc are not included. 
It has been shown by Pfalzner(2003, 2004a) that this
is a valid procedure as long as the disc is of low mass and one is
only interested in global features. Both conditions are fulfilled here,
as the discs chosen to be of mass $m_{disc}$=0.01 M$_{\sun}$ to correspond
to a value  characteristic for most observed discs,
and mass and angular momentum transfer are investigated. 

A description of the physical set up and the numerical method which is
used to simulate the encounters can be found in Section 2. From previous 
studies (\cite{heller:apj95,hall:mnras96,hall:mnras96}), the strongest impact
of the encounter on the disc is expected for prograde coplanar encounters. 
In Section 3 we start with this case and investigate how the mass and angular 
momentum transfer in the disc depends on the periastron distance
and the relative mass of the stars. We show how this transfer also depends
on the eccentricity of the encounter paths. The angular momentum 
change is investigated in two different ways - first we include all particles 
that are still bound to the star, then we include only the particles still
inside the initial disc size. The reason for using these two definitions is
the still unanswered question of how the disc matter looses 
sufficient angular momentum to be accreted onto its star.
Although probably collisions will not be the main mechanism for losing 
angular momentum, there is a certain amount of angular momentum loss due to 
encounters. As was pointed out by Pfalzner(2004), this angular momentum loss 
is usually underestimated by using this first definition. However, by
considering only particles inside the initial disc size a more physically
relevant angular momentum loss is found. 

In Section 4 a similar investigation for retrograde coplanar encounters 
is presented. Unless the encounter periastron is very close to the disc
``edge'', the mass and angular momentum transfer in such retrograde
encounters is much smaller than for prograde encounters of the same 
parameters. It has been argued that prograde encounters are more
likely than retrograde  but this is still an open question. The same
applies for whether the discs in a cluster are in anyway aligned.
Although there are some first studies (\cite{jensen:apj04}) concerning a
possible disc alignment, there is no definite answer yet.
Therefore in section 5 non-coplanar encounters are studied. It will be shown 
that even for inclinations of up to approximately 45 degree the mass and 
angular momentum loss  do not differ considerably from the prograde coplanar 
case. For larger inclinations  the effect on the disc is significantly 
diminished.
 
In section 6 the disc size after the encounter is studied
as function of the encounter parameters. Finally in section 7, it is 
determined how much disc matter is captured by the passing star.

\section{Disc model and numerics}

In this study both stars are allowed to move freely and only one star 
(hereafter called star 1) is surrounded by a disc.
Since observations indicate that low-mass discs are much more common than 
high-mass discs, in this study the disc mass is chosen as 
$m_{disc}= 0.01 M_{\sun}$ - a typical mass value of observed discs.
In this study the discs are simulated using 10 000 pseudoparticles as tracers 
of the observed gas. Earlier studies 
(Pfalzner 2003, 2004, Pfalzner et al. 2004) 
showed that in the case of such low-mass discs, the effects of 
pressure, viscous forces and self-gravity on the mass and angular momentum 
transport in such encounters are negligible. A test particle model 
therefore suffices, where the disc particles only feel the force of the 
two stars and vice-versa, however going beyond restricted 3-body simulations. 
Details of the numerical method are described in \cite{pfalzner:apj03}.
 
Despite the test model approach, these type of simulations are still
computationally expensive, so that higher resolution simulations can only be 
performed for particular cases. Benchmark simulations using 1 million 
particles revealed very little quantitative differences to simulations with 
10 000 particles indicating that the lower resolution suffices to determine 
the global features we are interested in here.  

In all simulations the discs extend initially from 10 to 100 AU, the gap of 
10 AU exists between the stars and the inner edges of the discs to avoid 
additional complex calculations of direct star/disc interactions and to save 
computer time. The density is distributed in the disc 
according to \[
\rho(r,z)=\rho_{0}(r)\,\exp\left(-\frac{z^{2}}{2H(r)^{2}}\right)\,,\]
where $H(r)$ is the local vertical half-thickness of the disc and is set constant in this study. $\rho_{0}(r)$ is the mid-plane density
with $\rho_0(r)\sim 1/r^2$ giving a surface density of $\Sigma\sim 1/r$,
where $\rho_0(r) = 1.313 \times$ 10$^{-10}$ g/cm$^3$ at $r$=1 AU. 
Star 1 is throughout the study of mass $M_{*}^1 = 1 M_{\sun}$ 
whereas the mass of star 2 varies in the different encounters between 
$M_{*}^2=0.1M_{\sun}$ and 2 $M_{\sun}$.
The periastra of the encounters are chosen between 100 and 450 AU.
As the  discs extends to $r_{disc}$ = 100 AU, this covers the parameter space 
from grazing to distant encounters. \\
As described in \cite{pfalzner:apj03} the obtained results can be 
generalized for other mass distribution within the disc by applying 
appropriate scaling factors. 

\section{Prograde coplanar encounters}

In this study we exclude the case where the encounter leads to the formation
of a bound system. Then of all possible encounter paths, parabolic encounters
lead to the longest interaction time and have the strongest impact on the 
disc. We will start our investigation by studying the effects of 
such parabolic prograde encounters. In this case the parameters which
characterize the encounters are the masses of the stars, $M_1^*$ and $M_2^*$,
and the periastron $r_{peri}$. In this study we choose to keep the
mass of star 1 constant at $M_1^*$ = 1 $M_{\sun}$  and vary $M_2^*$ and 
$r_{peri}$.

In Fig. \ref{fig:pro_para_peri} the 
dependence of the mass loss $\Delta m_{disc}/m_{disc}$ and angular momentum 
loss $\Delta J/J$ in the disc are shown as 
a function of $r_{peri}/r_{disc}$.  As expected, the mass and angular 
momentum transport decreases for larger 
periastra (see Fig. \ref{fig:pro_para_peri}). For the equal mass case  
$M_1^*=M_2^*=1 M_{{\sun}}$ the mass loss is approximately 48 \% and 
angular momentum loss $(\Delta J/J)_{total}$ about 62\% 
of the initial values for the closest investigated case of
($r_{peri}/r_{disc}$=1). For a lower mass of star 2  
($M_2^*=0.5 M_{\sun}$)  the mass loss is still 36 \% and total angular 
momentum about 46\%. For both cases an exponential decrease of the mass 
and angular momentum loss with $r_{peri}/r_{disc}$ is found.

For distant encounters Ostriker (1994) derived the following formula for
the angular momentum loss,
\be
\Delta J \sim \frac{M_2^*}{M_1^2+M_2^*} \exp 
\left[ -\sqrt{\frac{M_1^*}{M_1^2+M_2^*}} 
        \left( \frac{r_{peri}}{r_{disc}} \right)^{3/2}
\right]\nonumber \\
\times \frac{2}{\Omega(r_{disc})}\cos \left( \frac{\beta}{2}\right)^5,
\label{eq:ostriker}\ee 
where $\Omega(r_{disc})$ is the angular velocity at the outer disc radius and
$\beta$ the inclination between the disc plane and the orbital plane.
In the coplanar case considered here, $\beta$=0.
A comparison of this analytical result and our simulation data (see Fig.
\ref{fig:pro_para_peri}c) shows  good agreement for distant encounters, but 
for encounters closer than $r_{peri}/r_{disc}<$2 (in the case $M_2^*$ = 
1 $M_{\sun}$) the analytical results considerably overestimate the 
angular momentum loss.

Due to the encounter a portion of the disc 
particles are only loosely bound to the disc, undergoing eccentric orbits far 
outside the original disc size. These particles contribute to the total 
angular momentum but are irrelevant to the angular momentum loss
connected to accretion. Therefore in
Fig. \ref{fig:pro_para_peri}  and all following figures
that concern the angular momentum loss, the total angular momentum loss
$(\Delta J/J)_{total}$, which includes all particles bound to star 1,
as well as the angular momentum loss $(\Delta J/J)_{100}$ of the particles
bound within 100AU are shown. This is done to obtain a better measure of the
the angular momentum loss  {\it inside} the disc (Pfalzner 2004) 
relevant for the accretion process. As some of the bound particles move on 
strongly elliptical orbits, they leave and reenter the 100AU range 
periodically. This leads to a periodic change of $J_{100}$, but this change  
is in the considered parameter range below 1\%, so that it can be neglected.

As we can see already in Fig. \ref{fig:pro_para_peri}b  for certain encounter parameters $(\Delta J/J)_{100}$ can be 
considerably larger than $(\Delta J/J)_{total}$. 
$(\Delta J/J)_{100}$ decreases less steeply than $(\Delta J/J)_{total}$ for 
larger $r_{peri}/r_{disc}$.
In other words: for close encounters there is little difference between 
$(\Delta J/J)_{100}$ and $(\Delta J/J)_{total}$ whereas it can be considerable
for distant encounters. For example, for $M_2^*=1 M_{{\sun}}$
and $r_{peri}/r_{disc}$=2.75,  $(\Delta J/J)_{100}$ is more than twice
$(\Delta J/J)_{total}$. Since distant encounters are actually much more likely
to occur than close ones, this might be crucial for the loss of angular 
momentum to the star over time through repeated distant encounters
and might actually facilitate the accretion of matter from the disc onto the
star.
 
Fig. \ref{fig:pro_para_mass} shows how the mass and angular momentum loss
increases with the mass $M_2^*$ of star 2.
The mass and angular momentum loss are shown for different 
periastra, $r_{peri}$=100AU, 150AU and 200AU, which corresponds to  
$r_{peri}/r_{disc}$=1, 1.5 and 2 in Fig.\ref{fig:pro_para_peri}, respectively.
For small  masses of $M_2^*$, i.e. $M_2^*=0.1$, the losses in mass and angular 
momentum are all below 10\%  for all shown periastra, apart from
$(\Delta J/J)_{100}$ at 100AU. However, for higher masses $M_2^*$
the effect on the disc rises steeply but might eventually approach 
a certain maximum level for high masses $M_2^*$. For a periastron of 100 AU 
$\Delta m_{disc}/m_{disc}$ and  $(\Delta J/J)_{total}$ are around 55\%,
for $r_{peri}$=150AU they are about 35\% and for  $r_{peri}$=150AU still about
25\%. So that for a close encounter with low mass stars only a small effect
is expected, whereas a close encounter with a roughly equal mass star leads 
to the loss of a quarter to half the disc mass.  

The results in Figs. \ref{fig:pro_para_peri} and \ref{fig:pro_para_mass}
can be fitted by the following formula
\be
\frac{\Delta X}{X}=A \exp\left[-\frac{1}{2} 
\sqrt{\frac{M_1^*}{M_2^*}}\right]
\exp\left[-\sqrt{\frac{M_1^*}{M_2^*}}\left(\frac{r_{peri}-r_{in}}{r_{disc}}\right)^{3/2}\right].\nonumber
\\
\label{eq:mass}
\ee
For our results of the mass loss $\Delta X/X = \Delta m_{disc}/m_{disc}$  
best fitting is obtained with the parameters 
$A$ and  $r_{in}$ in Eq.\ref{eq:mass} being set $A=1.1$ and 
$r_{in}=40$ AU. Whereas for the total angular momentum loss
$\Delta X/X =(\Delta J/J)_{total}$ the data require $A=1.26$ and 
$r_{in}=70$ AU.

Eq. \ref{eq:mass} shows some similarities to the dependencies in the 
analytical result of the total angular momentum loss for distant encounters by 
Ostriker({1994).   
Comparing Eq. \ref{eq:mass} with the Ostriker result given by Eq. 1 
it can be seen that for the limit $r_{peri} \longrightarrow\ \infty$ 
the dependence on the periastron distance is recovered. However, the mass 
dependence differs. Ostriker derived Eq. \ref {eq:ostriker} under the
assumption of a small perturbation i.e. large periastron or small perturber
mass. For $ M_2^*>1 M_{\sun}$  a deviation from the analytical result is 
therfore not unexpected. For  $ M_2^*<1 M_{\sun}$  the deviation in the
results are probably due to the large statistical errors in the simulation
results.

The angular momentum loss within 100AU needs a slightly more complex fitting
formula
\be
\left(\frac{\Delta J}{J}\right)_{100}  = & A & \exp\left[-\frac{1}{2} 
\sqrt{\frac{M_1^*}{M_2^*}}\right]\nonumber \\
& \times &
\exp\left[-\sqrt{\frac{M_1^*}{M_2^*}\left(\frac{0.7(r_{peri}-r_{in})}{r_{disc}}\right)^{3}}\right]
\label{eq:mass}
\ee
Here $r_{in}$= 100AU but $A$ is different for each periastron, i.e. 
$A$=1 for $r_{peri}$=100AU,  $A$=0.85 for $r_{peri}$=150AU,  $A$=0.72 for 
$r_{peri}$=200AU.

Previous work (\cite{pfalzner:apj04}) suggests that in terms of mass and 
angular momentum transfer, hyperbolic encounters are ``failed'' parabolic 
encounters. The interaction time
in hyperbolic encounters is too short to enable the complete mass and angular 
momentum transfer that one would have in a parabolic encounter.
Fig. \ref{fig:pro_hyper} supports this view. It shows the mass and angular 
momentum loss as a function of the eccentricity of the encounter for
$M_2^*=0.5$ and $M_2^*=1.0$. 
The mass and angular momentum transfer values for the parabolic
encounter with $r_{peri}/r_{disc}$ = 100 AU is taken as reference and is 
therefore 1. Fig. \ref{fig:pro_hyper} shows how the mass and angular 
momentum loss in the encounter with the same mass of star 2 and periastron
becomes less efficient as the path becomes increasingly eccentric.
So for an eccentricity $\epsilon$=10 the mass and
total angular momentum transfer is only around 20\% of that in a parabolic 
encounter. As for the dependence on the periastron and the mass of the 
secondary star here again $(\Delta J/J)_{100}$ decreases less rapidly than 
$(\Delta J/J)_{total}$ and $\Delta m_{disc}/m_{disc}$ with increased
eccentricity.

The decrease in disc mass loss with increasing ellipticity seems
proceed in the same way for $M_2^*$=1$M_{{\sun}}$ as for 
$M_2^*$=0.5$M_{{\sun}}$. As both data point sets
can be fitted very well with an exponential decrease of 
$\exp(-0.4 (\epsilon-1))/\epsilon^{0.5}$. However, it should be noted that
although the shape of the curve 
for $(\Delta J/J)_{100}$ and $(\Delta J/J)_{total}$ is for both masses
very similar the $M_2^*=0.5$ values are always a bit less than 
$M_2^*=1.0$ values.

A mention about the errors is in order. There are several sources of error; 
a) the limited resolution of the disc, b) deviations from the ideal parabolic 
path and c) for close encounters the actual particle distribution at the 
``edge'' of the disc might play a role. Comparing the results of different
simulations with the same interaction parameters, we conclude that for
prograde encounters the absolute error of the mass and angular momentum loss 
is about 2-3\%.

\section{Retrograde coplanar encounters}

Retrograde encounters have been investigated to a much lesser degree than
prograde encounters. There are two reasons for this, first, the few previous
results (\cite{heller:apj95,hall:mnras97}) indicate that retrograde encounters 
change the mass and angular momentum distribution 
in the disc far less, second, it has been argued that prograde 
encounters are more likely to occur than retrograde encounters. The
second reason is based on the assumption that star-disc
systems that encounter each other are likely to have a common formation 
history. If this is the case, there could exist a preference for the discs
of having the same rotational orientation as well. 
Obviously, this argument is strongly linked to the underlying star formation
picture. To our knowledge there exists no observational evidence that would
suggest a preference of prograde orientations of neighbouring disc systems.
As long as this point is not clarified, the same information about 
the effect of encounters on the discs is needed  retrograde encounters
as for the prograde case. 

In the analytical result of Ostriker in Eq. \ref{eq:ostriker} the retrograde 
case is equivalent to $\beta=\pi$. In this case Eq. \ref{eq:ostriker} 
predicts that there is no angular momentum loss. Although this is valid for 
distant encounters, our simulation results show that this is not true for 
close encounters.

Fig. \ref{fig:retro_para_peri} illustrates the dependence of the mass and 
angular momentum loss in the disc as function of $r_{peri}/r_{disc}$ for the
retrograde case in parabolic encounters. In agreement with previous results 
(\cite{heller:apj95,hall:mnras97}) there is no mass and total angular 
momentum loss for encounters more distant than $r_{peri}/r_{disc}$=1.5. 
It can be seen that even in the periastron range of  
1.25  $ < r_{peri}/r_{disc} <$  1.5 the change in mass and total angular 
momentum is always less than 3\%. However, there is a steep increase in 
$(\Delta J/J)_{total}$ and $\Delta m_{disc}/m_{disc}$ as the encounters
become so close that the passing star nearly penetrates the disc. For example,
in the case $M_2^*=1.0$ and $r_{peri}/r_{disc}$=1, the mass as well as
$(\Delta J/J)_{total}$ is about 13-14\%. However,
even in these cases the mass and angular momentum loss in retrograde 
encounters are about a factor 3 less than for the same prograde encounter.

Here again $(\Delta J/J)_{100}$ is a much less steep function of
$r_{peri}/r_{disc}$ than $(\Delta J/J)_{total}$ and 
$\Delta m_{disc}/m_{disc}$. Interestingly  so that for example at 
$r_{peri}/r_{disc}$=1.5 and $M_2^*$=1$M_{{\sun}}$ there is still about 
8\% loss of $(\Delta J/J)_{total}$, but non for the mass or the total 
angular momentum.

% figure for the angular momentum distribution inside disc?

Fig. \ref{fig:retro_para_mass} shows how the mass and the total angular 
momentum loss depends on $M_2^*$ for a retrograde encounter. As only for
very close encounters considerable mass and angular momentum transport can 
be expected, the cases $r_{peri}$=100AU and $r_{peri}$=110AU were 
investigated. The shape of the curves is similar to the prograde case
shown in Fig.\ref{fig:pro_para_mass}, which points to a similar depends
on  $M_2^*$ as in the prograde encounter. However, the values of 
$\Delta m_{disc}/m_{disc}$ and $\Delta J/J$ are much less.
For example, is   $\Delta m_{disc}/m_{disc}$ for $r_{peri}/r_{disc}$=1.0
and $M_2^*$=1.5$M_{{\sun}}$ for this retrograde encounter less than half 
of that in the prograde encounter.

How do $\Delta m_{disc}/m_{disc}$ and $\Delta J/J$ depend on the 
eccentricity of the encounter in the retrograde case? 
Fig. \ref{fig:retro_ell} shows that the mass and angular
momentum loss is much less even if the encounter is only slightly 
hyperbolic. So even slight deviations from the parabolic case seem to 
affect retrograde encounters more than parabolic encounters. 
This means there is a larger error to be expected for the
retrograde data than for the prograde results.

Apart from the prograde parabolic nearly penetrating case,
retrograde encounters cause no mass transfer or changes 
to $J_{total}$ occur. Only $J_{100}$ seems to be effected.

\section{Non-coplanar encounters}

What happens if the encounter is not coplanar?
From previous studies it is known that although non-coplanar encounters
can lead to warping etc., less mass and angular momentum are lost
than in prograde coplanar encounters. Here we attempt to quantify the effect 
of non-coplanarity. We investigate 4 different cases given by 
parameter sets $r_{peri}/r_{disc}$ and $M_2^*$, varying the angle 
of inclination between the plane of the path of star 2 and the disc 
orientation. The result is shown in Fig.\ref{fig:non}. Here the coplanar 
prograde encounter corresponds to an angle of 0 degree and the coplanar 
retrograde encounter to an angle of 180 degree.
  
Fig.\ref{fig:non} shows that there is the general trend for all 4 cases 
that a slight deviation from coplanarity does not effect 
$\Delta m_{disc}/m_{disc}$ or $\Delta J/J$ very much,
only for inclinations larger than 45 degree this changes.

We now look at the 4 different sets separately.
These 4 parameter sets were chosen in such a way that they represent 
different typical situations. It is necessary to investigate a distant 
encounter to be able to compare with the analytical treatment of 
distant encounters by \cite{ostriker:apj94}. The case $M_2^*$=1.25 $M_{\sun}$ 
and 
$r_{peri}/r_{disc}$=2 represents such a distant encounter, where the
relatively high mass of star 2 ensures that in the prograde coplanar case the 
losses are still larger than a few percent to be able to clearly distinguish
them from the errors which are typically in the 2-3\% range.

The corresponding retrograde encounter leads to no mass or angular momentum
transfer.  Fig.\ref{fig:non} shows that there is a smooth transition from the
coplanar encounter value to this  retrograde value. For more than
approximately 100-120 degree of inclination there is no more change in
$\Delta m_{disc}/m_{disc}$ or $(\Delta J/J)_{total}$. For $(\Delta J/J)_{100}$
there is some change at all angles, which is nevertheless
highest for coplanar encounters.

The case $M_2^*$=0.75 and $r_{peri}$=150AU shows that
although a closer encounter more or less the same dependence on the
relative orientation of the path and the disc. 
Only the absolute values for $\Delta m_{disc}/m_{disc}$ and $\Delta J/J$
are larger.

The situation is different for $M_2^*$=1.25 and $r_{peri}$=200AU,
here in the retrograde encounter, mass and angular momentum are lost as well.
The data seem to indicate that the minimum of mass and angular momentum 
transfer is not for the retrograde coplanar case but somewhere between
120 and 180 degrees. As the loss values are very small, this could be
just a statistical error. 

However, the case $M_2^*$=1.25 $M_{\sun}$ and $r_{peri}$=100AU leaves
no room for doubt. Here the retrograde value is well out of this error
range and here again the smallest transfer of mass and angular momentum
lies between 120 and 165 degree. So that the least mass and angular
momentum transfer can be expected for slightly non-coplanar retrograde 
encounters.  
 
In Fig. \ref{fig:non}c we test whether in the non-coplanar case
$(\Delta J/J)_{total}$ shows the $cos(\beta/2)^5$-dependence of the
Ostriker result (Eq.1) on the inclination between the orbital and the disc 
plane. This was done for the two cases  $r_{peri}/r_{disc}$=2, $M_2^*$=1.25
and $r_{peri}/r_{disc}$=1, $M_2^*$=0.75. In the more distant case the
simulation results agree reasonable well with this $cos(\beta/2)^5$-dependence.However, for closer encounters the dependence on the inclination angle is
much more complex.

\section{Disc mass transport between stars}

In encounters mass can not only be lost from the disc and drift away from
star 1, but star 2 can also capture disc matter.  
An interesting situation arises if one studies encounters where
both stars are surrounded by discs. Then the mass transported between the 
discs can lead
to chemical mixing of the two disc components.
\cite{pfalzner:05} showed that the results for star-disc encounters
can be directly applied to disc-disc encounters, as long as the disc masses
are low. In the following therefore this capturing of disc material by 
star 2 will be investigated as well with the chemical mixing in disc-disc 
encounters in mind. 

From our simulations we find that 
such mass capture of disc material nearly exclusively happens
for prograde encounters. The only exception is if star 2 nearly or actually
penetrates the disc. So in the following we restrict the investigation to
prograde, coplanar parabolic encounters again. 

In  Fig. \ref{fig:capture} the relative disc mass captured by star 2 in 
such prograde, coplanar parabolic encounters is shown. It can be seen 
how the captured mass depends on a) the mass $M_2^*$  of star 2 and b) the 
relative periastron distance $r_{peri}/r_{disc}$ in the encounter. For low 
masses of star 2 ($M_2^*$=0.1 M$_{\sun}$) the captured mass is below 5\% of 
the disc mass even for grazing encounters with $r_{peri}$=100AU. However, for
higher masses of star 2 ($M_2^*$=1.5 M$_{\sun}$) up to 20\% of the mass of 
the disc around star 1 can be captured by star 2. For close encounters 
($r_{peri}/r_{disc}$=1.0 and 1.5) the captured mass is a steep function of 
the mass of star 2 as long
as $M_2^* <$ 0.5 M$_{\sun}$, for higher masses ($M_2^* >$ 0.8 M$_{\sun}$)
the captured mass seems to depend much less strongly on $M_2^*$. The curves in 
Fig. \ref{fig:capture}a become rather flat for $M_2^* > $0.8 M$_{\sun}$
indicating  that there is possibly an upper limit for the mass transfer to 
the second star for each periastron distance.

Fig. \ref{fig:capture}b shows the dependence
of the captured mass on $r_{peri}/r_{disc}$ for $M_2^*$=0.5 M$_{\sun}$
and $M_2^*$=1 M$_{\sun}$. The decrease in captured mass for larger
$r_{peri}/r_{disc}$ seems to depend nearly linear on the periastron distance, 
with a similar slope for  $M_2^*$=0.5 M$_{\sun}$ and $M_2^*$=1 M$_{\sun}$. 
For encounters more distant than 200AU the captured mass is well below 10\%.

\section{Disc size}

The typical effect in such encounter is that some mass is lost
from the outside of the discs as well some mass transported inwards. 
The density profile in the disc is changed and due to the relative 
low density at the outside of the disc the whole disc will appear smaller.  
 
An important question for the likelihood of the formation of planetary systems
is the survival rate of discs in a cluster. The collision rate as well as the 
effect of an encounter on the disc size are important parameters in this 
context. Here we want to investigate the latter and see how the disc 
size after the encounter depends on the encounter parameters. The
study will be limited to prograde coplanar, parabolic encounters for the
already above described reason that retrograde encounters influence the disc 
to a much lesser degree. In addition in the case of non-coplanar encounters, 
discs become warped which would make the definition of disc size more 
complex.

In section 2 it was shown that the initial disc has a sharp cut-off
at 100AU. In reality such a sharp cut-off does not exist but a smooth
transition towards lower disc densities for larger distances. In the 
following the disc size is not defined by this 100AU limit, but as the radius 
were 95\% of the disc mass are enclosed. For our disc that is at 92AU. 

To this the effect of an encounter Fig. \ref{fig:discsize} shows the 
relative disc size (with the unperturbed disc as reference value 1)
after encounters with different parameters.
Fig. \ref{fig:discsize}a demonstrates that the disc size remains
relative unaltered for distant encounters (here indicated 
by the black triangle for the $r_{peri}$=200AU data). Even
for relatively high masses of star 2 ($M_2^*$=1.5 M$_{\sun}$) the disc size
is still 90\% of the initial disc size. 

However, for grazing encounters (indicated by $\bullet$ for $r_{peri}$=100AU) 
the disc 
size can be reduced by up to 40\%, if the mass of star 2 is large enough 
($M_2^*$=1.5 M$_{\sun}$). However, if star 2 is of much lower mass
than star 1 (M$_*^2<$ 0.4M$_*^1$ ), even in grazing encounters the
disc size is not reduced to less than 90\% of its original size.  
 
Fig. \ref{fig:discsize}b supports these finding. Here the relative disc size
is shown as a function of $r_{peri}/r_{disc}$. Again it can be seen
that the disc size is reduced by less than 10\% for distant encounters 
($r_{peri}/r_{disc} >$ 2). Only for closer encounters the disc size is 
considerably reduced. 

From these considerations it can be concluded that relatively close 
encounters with a high mass of star 2 are necessary to really reduce the 
disc size. The effect on the disc is much more an alteration of mass 
distribution profile than the disc radius as such.

\section{Summary}

In this paper we presented a first parameter study of the effect
of encounters between two stars where one is surrounded by a 
low-mass protoplanetary accretion disc for all major types of 
encounters, i.e. prograde/retrograde; coplanar/non-coplanar and 
parabolic/hyperbolic.  The mass loss of the disc, its angular 
momentum change, the resulting disc size and the amount of matter captured by 
the star were investigated. 

As indicated by the special cases studied in previous work, this
parameter study confirms that coplanar, prograde parabolic encounters lead 
to the highest mass and angular momentum loss; for example nearly 
50\% of the initial disc mass and around 60\% of the
angular momentum are lost in an
equal stellar mass encounters with $r_{peri}/r_{disc}$=1.
Retrograde encounters lead to very little mass and angular momentum loss
in the disc unless they are nearly grazing. Even for grazing encounters
only high mass perturbers have a significant effect on the disc.
For the prograde and retrograde case parabolic encounters are
naturally most disruptive to the disc whereas hyperbolic encounters due
to there shorter interaction time have less effect on the disc.
Retrograde encounters seem to be more sensitive to deviations from the parabolic
path, as even slightly non-parabolic encounters lead to significantly
less mass and angular momentum loss.

In the case of non-coplanar encounters, up to an inclination of 45 degree
between the path of star 2 and the disc plain of star 1 mass and angular
momentum losses are slightly less than in the coplanar prograde case.
However, for more inclined systems the mass and angular momentum loss
drop significantly; for most encounters to zero, only for nearly grazing 
encounters there can be mass and angular momentum transport at such 
inclinations. For latter the minimum of mass and angular transport is not for
the retrograde (180 degree) case, but somewhere between 120 and 165 degree.

As far as possible fitting formulae for most of the above results are given.
So that a much wider parameter space than previously is accessible.

In addition the capturing of mass by star 2 was investigated.

Our aim is to combine these results with simulations of dense clusters
to be able to make predictions of the lifetime of disc in dense stellar
clusters.

\begin{figure}
%\resizebox{\hsize}{!}{\includegraphics{para_pro.eps}}
\resizebox{\hsize}{!}{\includegraphics{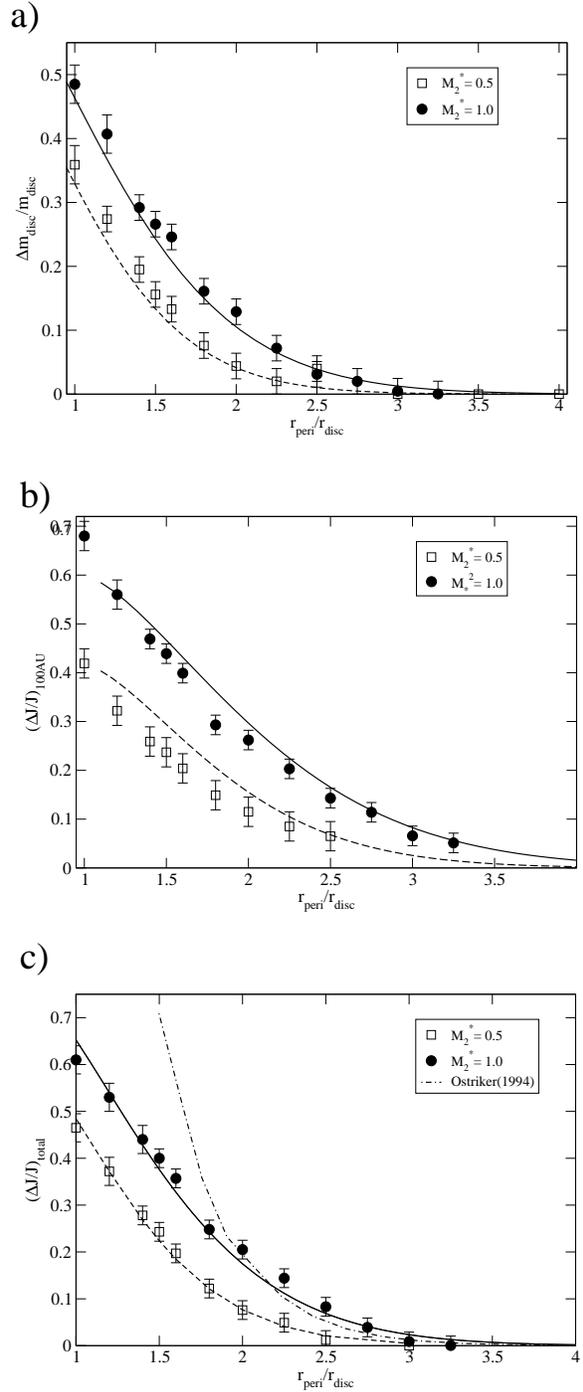}}
\caption{Prograde encounters: Relative  a) mass loss and  
b) angular momentum loss within 100AU  and c) in total as a function of the 
periastron from the primary star. The periastron is given in units of the 
disc size, in the here considered case that was 100AU. $M_2^*$ is given in 
units of solar masses. The analytical result of the total angular momentum
loss by Ostriker(1994) is indicated in c).}
\label{fig:pro_para_peri}
\end{figure}

\begin{figure}
%\resizebox{\hsize}{!}{\includegraphics{para_pro_mass2.eps}}
\resizebox{\hsize}{!}{\includegraphics{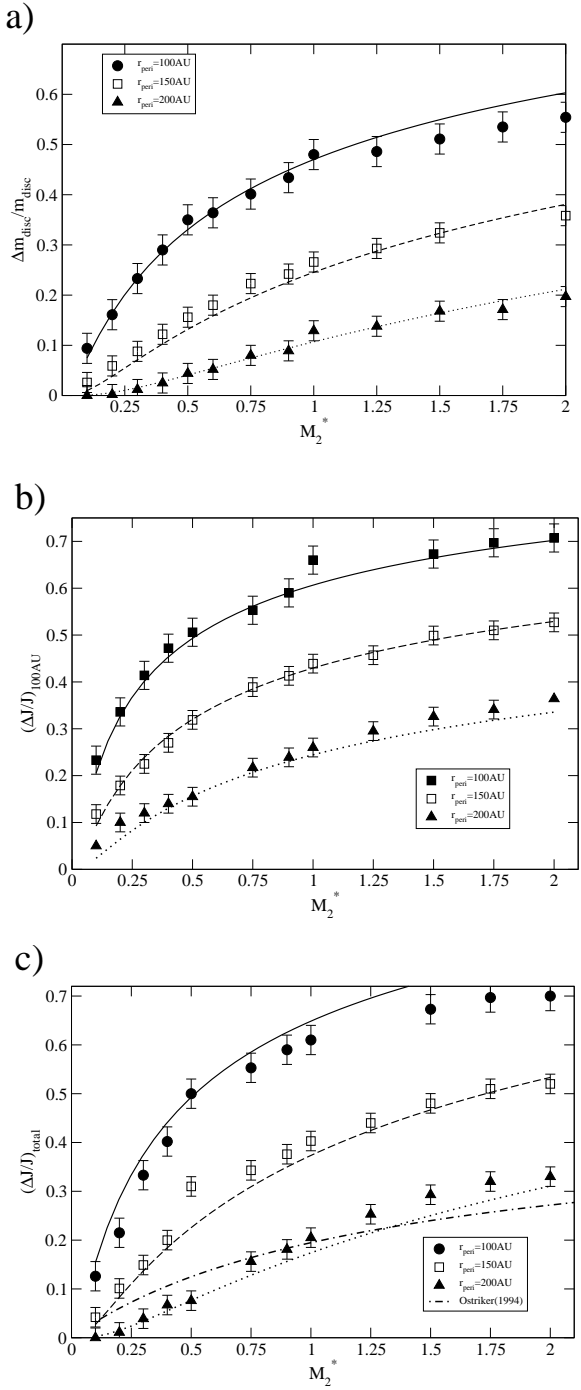}}
\caption{Prograde encounters: Relative  a) mass loss and  b) angular 
momentum loss b) within 100AU  and c) in total as a function of the mass 
$M_2^*$ (in units of solar masses) of star 2 for encounters with periastra of 100, 150 and 
200 AU. In c) the result of Ostriker (1994) is shown in comparison for the case
$r_{peri}/r_{disc}$=200AU.}
\label{fig:pro_para_mass}
\end{figure}

%\begin{figure}
%\epsscale{1.00}
%\plotone{prograde.eps}
%\caption{Loss of a) disc mass, b) angular momentum in the entire disc
%and c) angular momentum loss within the size of the original disc in per cent
%as function of i) the mass of the secondary, ii) the periastron and iii)
%the velocity in the periastron.
%\label{fig:pro_mpv}}
%\end{figure}

%\begin{figure}
%\epsscale{0.55}
%\plotone{retro.eps}
%\caption{Change of the angular momentum  a) per 
%particle and b) in total as a function of the radial distance from
%the primary star.
%\label{fig:ang_rad_simple}}
%\end{figure}

\begin{figure}
%\resizebox{\hsize}{!}{\includegraphics{para_ell.eps}}
\resizebox{\hsize}{!}{\includegraphics{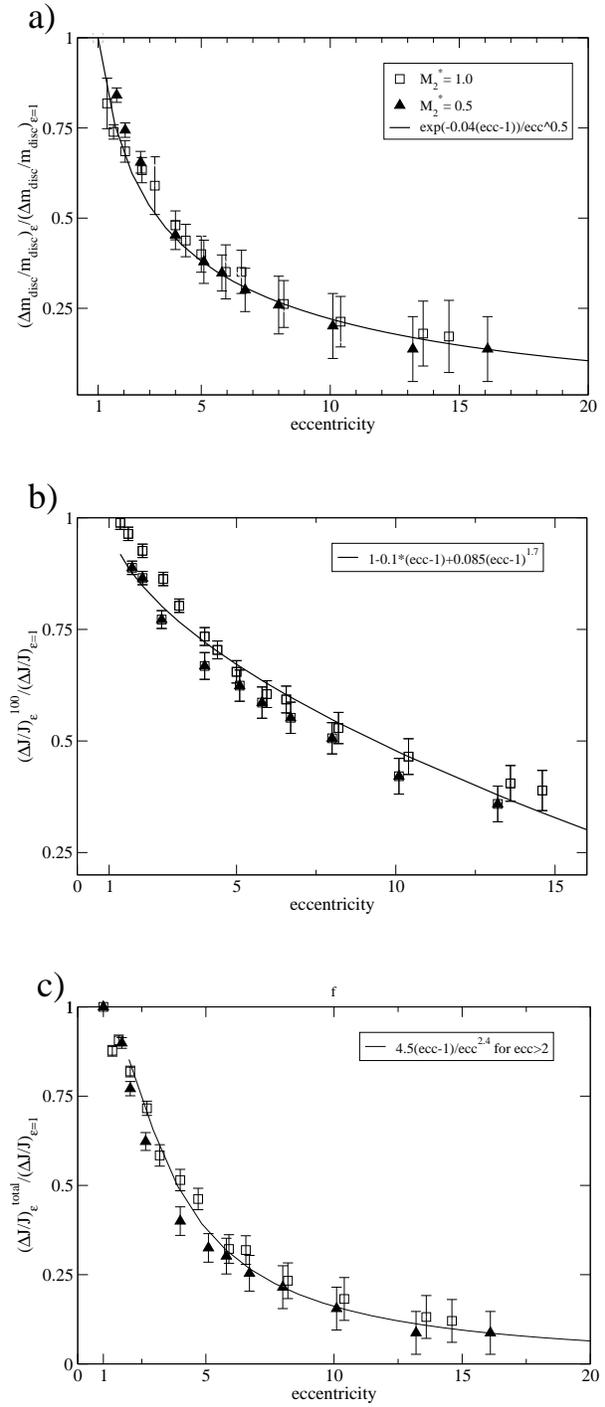}}
\caption{Prograde encounters: Relative  a) mass loss and  b) angular 
momentum loss within 100AU  and c) in total as a function of the eccentricity of the orbit of star 2. The values for the parabolic encounter are
chosen as points of reference(=1).}
\label{fig:pro_hyper}
\end{figure}

\begin{figure}
%\resizebox{\hsize}{!}{\includegraphics{retro_peri.eps}}
\resizebox{\hsize}{!}{\includegraphics{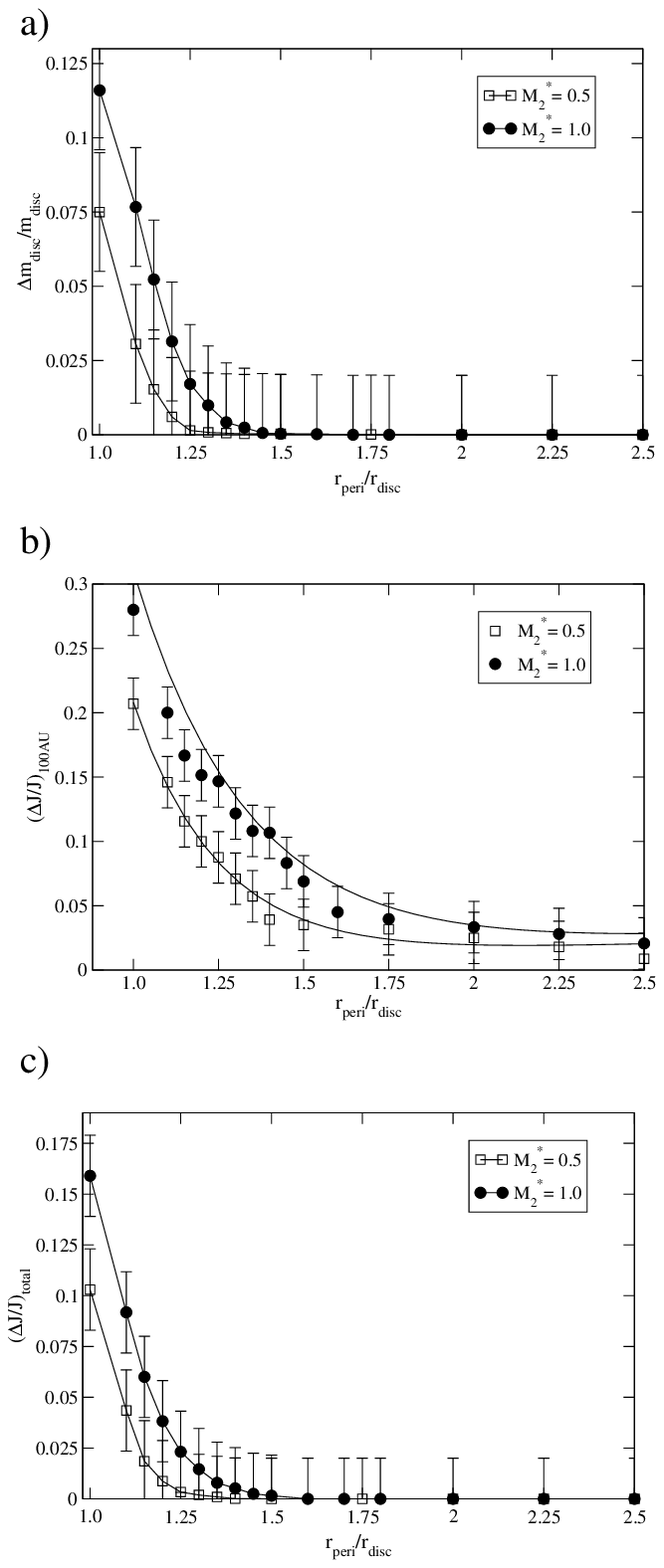}}
\caption{Retrograde encounters: Relative  a) mass loss and  
b) angular momentum loss within 100AU  and c) in total as a function of the 
periastron from the primary star. The periastron is given in units of the 
disc size, in the here considered case that was 100AU.$M_2^*$ is given in units of solar masses.}
\label{fig:retro_para_peri}
\end{figure}

\begin{figure}
%\resizebox{\hsize}{!}{\includegraphics{retro_pro_mass2.eps}}
\resizebox{\hsize}{!}{\includegraphics{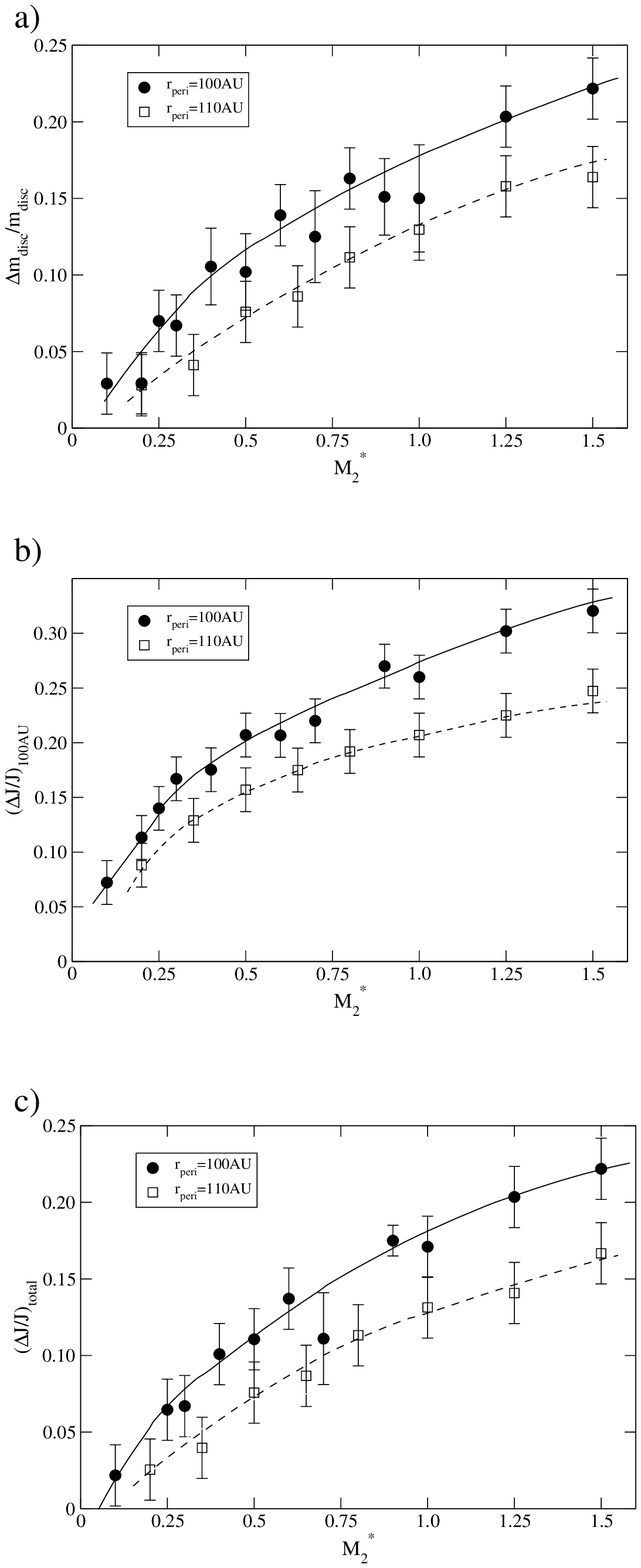}}
\caption{Retrograde encounters: Relative  a) mass loss and  b) angular 
momentum loss b) within 100AU  and c) in total as a function of the mass 
$M_2^*$ of the secondary star for encounters with periastra of 100 and 
110 AU. $M_2^*$ is given in units of solar masses.}
\label{fig:retro_para_mass}
\end{figure}

\begin{figure}
%\resizebox{\hsize}{!}{\includegraphics{retro_ell.eps}}
\resizebox{\hsize}{!}{\includegraphics{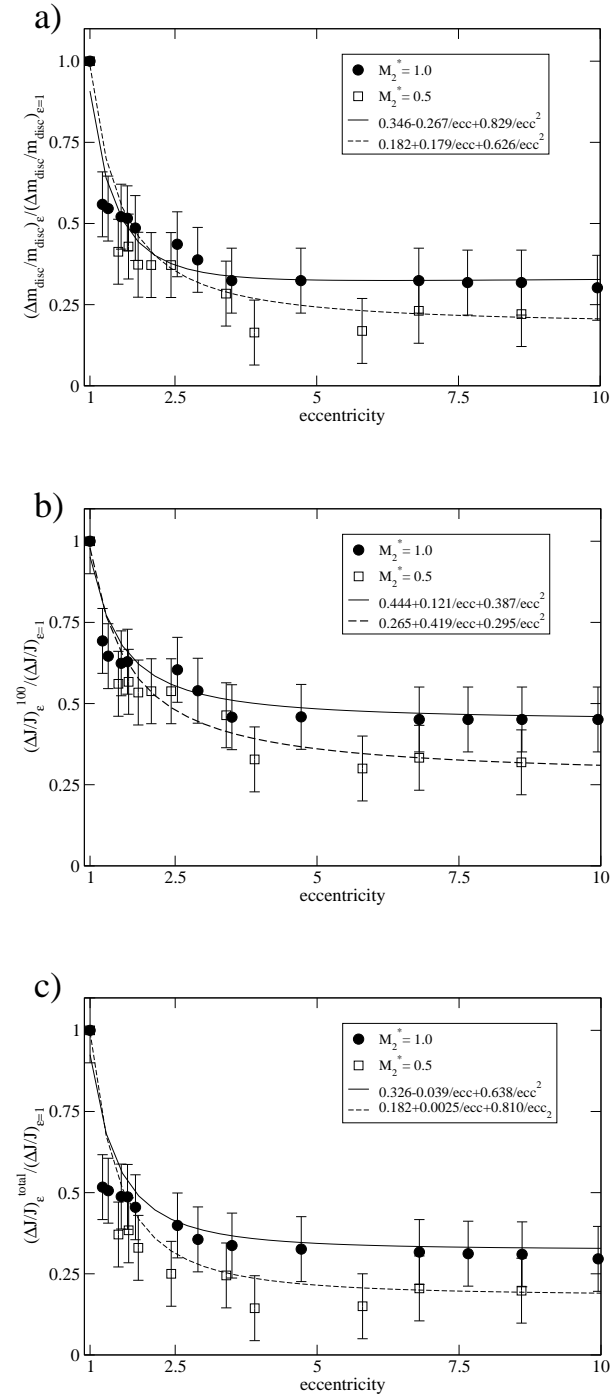}}
\caption{Retrograde encounters: Relative  a) mass loss and  b) angular 
momentum loss b) within 100AU  and c) in total as a function of the 
eccentricity of the encounter. $M_2^*$ is given in units of solar masses.}
\label{fig:retro_ell}
\end{figure}

\begin{figure}
%\resizebox{\hsize}{!}{\includegraphics{non.eps}}
\resizebox{\hsize}{!}{\includegraphics{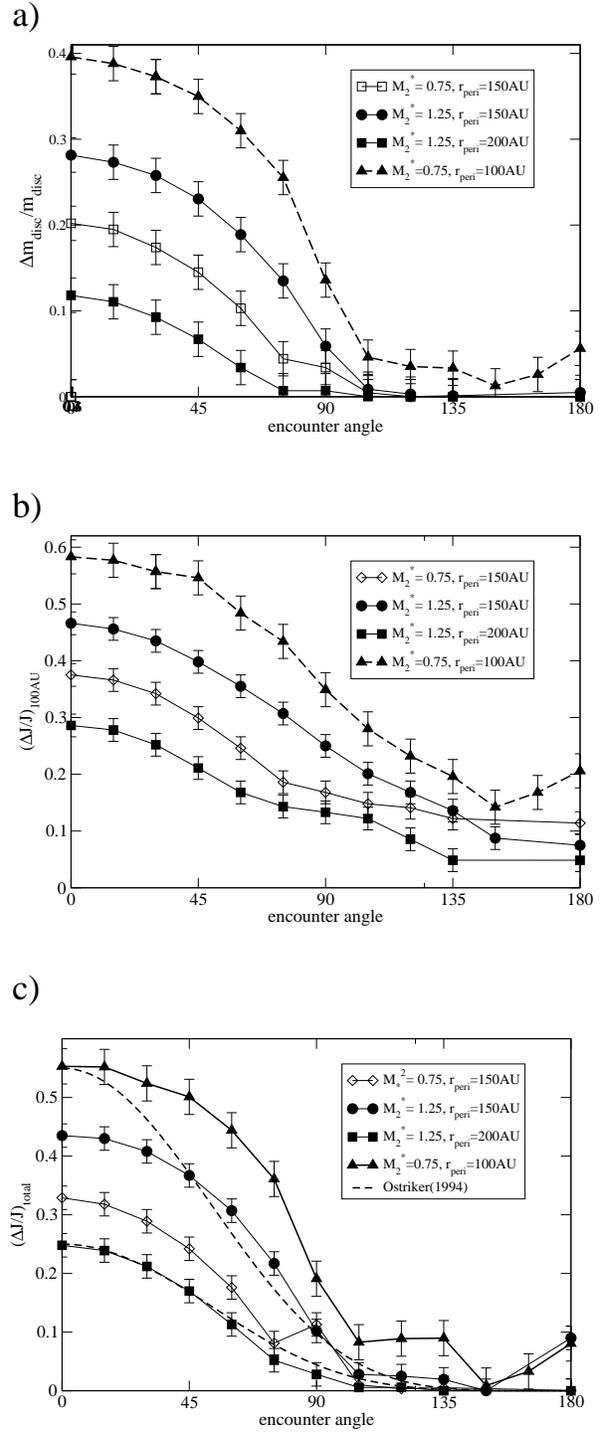}}
\caption{Non-coplanar encounters: Relative  a) mass loss and  b) angular 
momentum loss b) within 100AU  and c) in total as a function of the mass 
$M_2^*$ of the secondary star for encounters with periastra of 100, 150 and 
200 AU. The analytical result of the total angular momentum
loss by Ostriker(1994) is indicated in c) for most distant encounter
with $r_{peri}$=200AU and the strongest encounter with $r_{peri}$=100AU
and $M_2^*=0.75 M_{\sun}$.}
\label{fig:non}
\end{figure}

\begin{figure}
%\resizebox{\hsize}{!}{\includegraphics{capture.eps}}
\resizebox{\hsize}{!}{\includegraphics{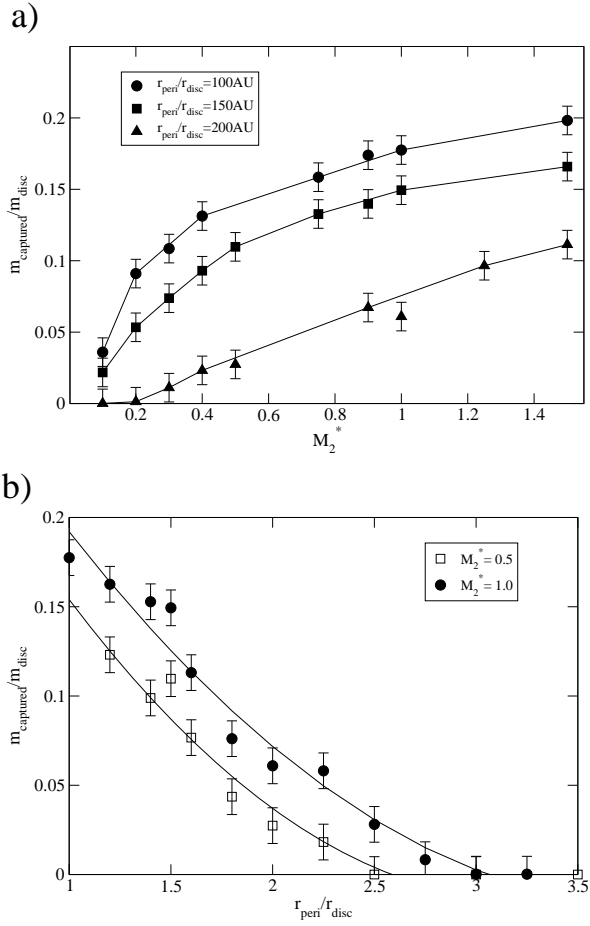}}
\caption{Relative mass captured from the disc of star 1 by star 2 as
function of a) the mass of star 2, b) the relative periastron 
$r_{peri}/r_{disc}$. In a) the simulations were performed for 
$r_{peri}$= 100 AU, 150 AU and 200 AU. In b) the cases
of $M_2^*$= 0.5 $M_{\sun}$ and $M_2^*$= 1 $M_{\sun}$ were investigated.
$M_2^*$ is given in units of solar masses. }
\label{fig:capture}
\end{figure}

\begin{figure}
%\resizebox{\hsize}{!}{\includegraphics{discsize.eps}}
\resizebox{\hsize}{!}{\includegraphics{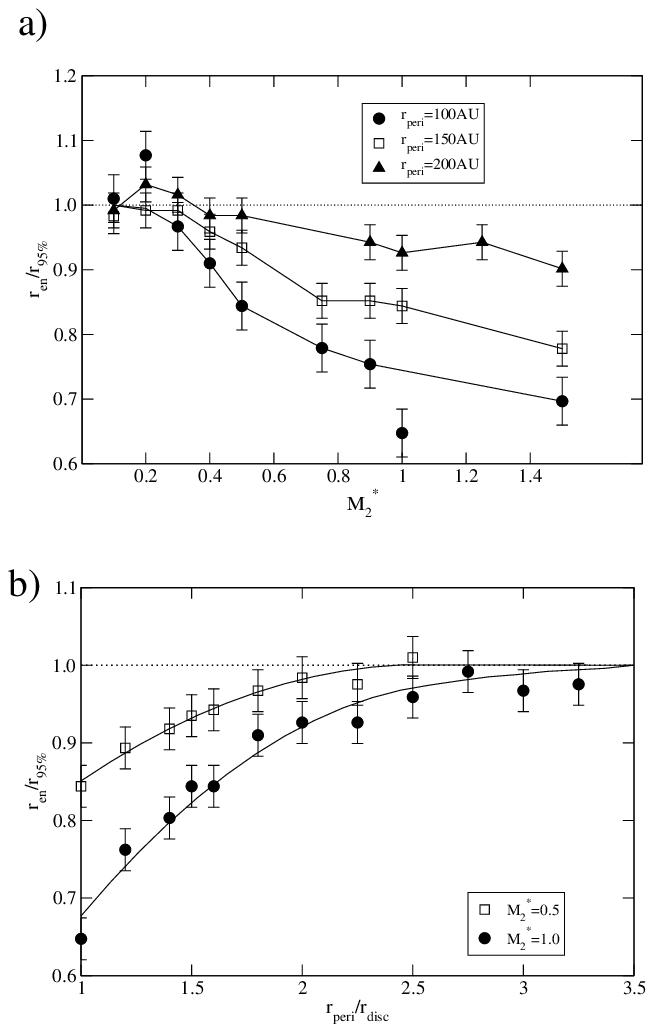}}
\caption{Relative disc size after the encounter as a function of a) the mass 
of star 2, b) the relative periastron $r_{peri}/r_{initial}$. In a) 
the simulations were performed for $r_{peri}$= 100 AU, 150 AU and 200AU. 
In b) the cases of $M_2^*$= 0.5 $M_{\sun}$ and $M_2^*$= 1 
$M_{\sun}$ were investigated. $M_2^*$ is given in units of solar masses.}
\label{fig:discsize}
\end{figure}

\end{document}